\documentclass[lettersize, journal]{IEEEtran}		

\usepackage{amsmath,amsfonts}
\usepackage{algorithmic}
\usepackage{adjustbox}
\usepackage{caption}
\usepackage{subcaption}
\usepackage{array}
\usepackage[caption=false,font=normalsize,labelfont=sf,textfont=sf]{subfig}
\usepackage{textcomp}
\usepackage{stfloats}
\usepackage{url}
\usepackage{authblk}

\usepackage{verbatim}
\usepackage{graphicx}

\usepackage[table]{xcolor} 
\usepackage{tabularray}
\usepackage{tabularx}

\usepackage{cite}

\usepackage{multicol}
\usepackage{multirow}
\usepackage{balance}	

\usepackage[switch]{lineno} 

\def\BibTeX{{\rm B\kern-.05em{\sc i\kern-.025em b}\kern-.08em
    T\kern-.1667em\lower.7ex\hbox{E}\kern-.125emX}}

\newcolumntype{s}{>{\hsize=.33\hsize\linewidth=\hsize}X}
\newcolumntype{D}{>{\hsize=.4\hsize\linewidth=\hsize}X} 

\newcommand{\wdImg}{\dimexpr \linewidth} 

\begin{document}

\title{Requirements on bit resolution in optical Ising machine implementations 

\thanks{\hrule \vspace{0.25em}  *Correspondence and requests should be addressed to T.S. (email: toon.jan.b.sevenants@vub.be) or to G.V. (email: guy.verschaffelt@vub.be)}}

\author[*1]{T. Sevenants}
\author[1]{G. Van der Sande}
\author[1]{G. Verschaffelt}

\affil[1]{Applied Physics Research Group, Vrije Universiteit Brussel, Pleinlaan 2, 1050 Brussels, Belgium}

\maketitle

\begin{abstract}
\noindent Optical Ising machines have emerged as a promising dynamical hardware solver for computational hard optimization problems. These Ising machines typically require an optical modulator to represent the analog spin variables of these problems. However, modern day optical modulators have a relatively low modulation resolution. We therefore investigate how the low bit-resolution of optical hardware influences the performance of this type of novel computing platform. 
Based on numerical simulations, we determine the minimum required bit-resolution of an optical Ising machine for different benchmark problems of different sizes. Our study shows that a limited bit-resolution of 8bit is sufficient for the optical modulator. Surprisingly, we also observe that the use of a 1bit-resolution modulator significantly improves the performance of the Ising machine across all considered benchmark problems.
\end{abstract}

\section{Introduction} \label{sec: Introduction}


\IEEEPARstart{I}{n}  recent years, there has been a growing interest in novel computing paradigms that could replace current traditional digital computer systems in applications that require high computational power and consume large amounts of energy, such as solving combinatorial optimization problems \cite{Anand2020}. It is believed that today's traditional digital hardware will struggle to find the solution to these hard optimization problems in a time- and energy-efficient manner. The underlying reason stems from their inherent limitations. A first argument for this statement is the Von Neumann bottleneck that is present in almost all digital hardware. This bottleneck arises because the central processing unit (CPU) must communicate with the memory unit, which becomes a time- and energy-inefficient process in these high-performance applications. Secondly, over the last years, Moore's Law has been observed to be stagnating, meaning that the increase in computing power of digital systems is also expected to slow down \cite{Horowitz2014}. 

 At the same time, there exist numerous important societal optimization problems that demand substantial computational power. Examples of such optimization problems can be found in chip design \cite{Barahona1988}, route planning \cite{Lucas2014},  financial portfolio management \cite{Horvath2016}, or even cryptography \cite{Saad2001, Wang2020}. The Ising machine (IM) has emerged as a new and promising computing platform that could solve these hard optimization problems in a time- and energy-efficient manner \cite{Wang2013, Lucas2014, Haribara2016, Yamamoto2017, Mohseni2022, kalinin2022}. 	

The general working principle of an IM is not based around the Von Neumann architecture, but rather around the mathematical concept that it is possible to map the cost function of a combinatorial optimization problem onto the Ising Hamiltonian:
\begin{equation}
\label{eq: ising_energy}
    H_{ising} = -\frac{1}{2} \sum_{ij} J_{ij} \sigma_i \sigma_j
\end{equation}
where $\sigma_i = \{-1, 1\}$ are binary variables, which are referred to as spins by convention, and $J_{ij}$ is the interconnection matrix that tells us which spins are connected and by which weight \cite{Lucas2014, Haribara2016}. The IM is a physical implementation of this binary interconnected network and like any dissipative physical system, the IM will have the natural tendency to minimize its energy, which it can do by flipping its spins. Due to the mapping of the problem's cost function onto the Ising Hamiltonian, finding the ground-state spin configuration corresponds to finding the optimum solution of the encoded optimization problem. Different problems have different cost functions, i.e. different energy functions, to be minimized, which is expressed as different interconnection matrices \textbf{J}.

\par
\begin{figure}[!t]
\centering
\includegraphics[width=1\columnwidth]{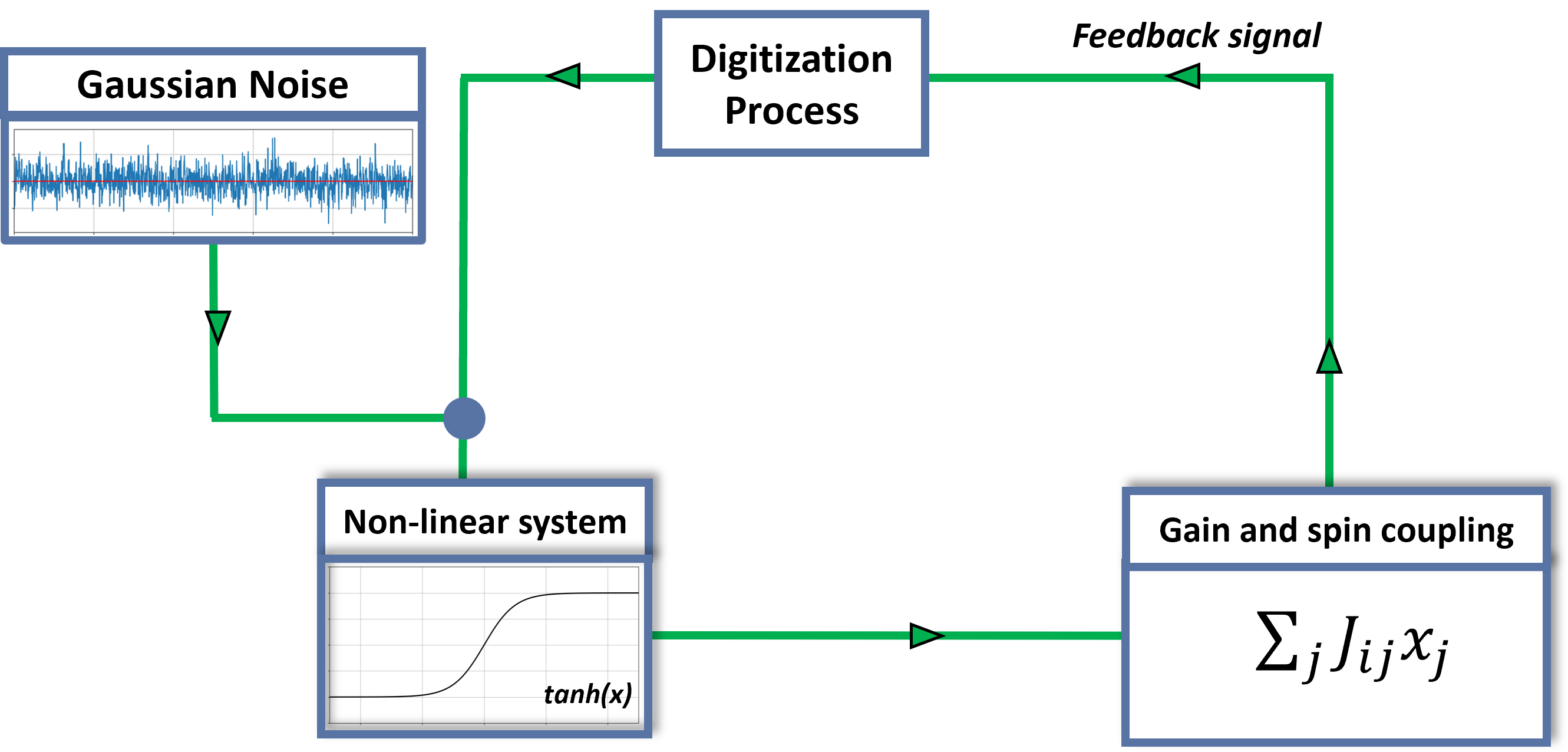} 
\caption{ Essential building blocks of an analog Ising machine, where the specific type of non-linearity depends on the hardware being used.}
\label{fig: network_example}
\end{figure}

There already exist multiple hardware implementations of the IM \cite{Yamamoto2017, Berloff2017, Shim2017, Pierangeli2019, Fabian2019,  Honjo2021, Mohseni2022, English2022, Litvinenko2023}, which are all built around three major elements: a type of non-linearity to implement the spin-variables, a form of spin-coupling to implement the connection matrix, and finally a feedback signal that is transmitted back to the non-linearity. The schematic layout of such IMs is illustrated in Fig.~\ref{fig: network_example}. In this paper, we will mainly focus on optical hardware implementations as they potentially offer several advantages over electronic implementations, such as higher bandwidth, lower latency, lower energy consumption and massive spatial multiplexing \cite{Spall2020, Stark2020, Margalit2021, Wang2022}. 

\par
A first example of such an optical hardware implementation is the Coherent Ising machine (CIM) which uses degenerate optical-parametric oscillator (DOPO) pulses to encode the spin state of the Ising network and an FPGA to facilitate the spin-coupling and calculate the spin-feedback \cite{Wang2013, Yamamoto2017, Inagaki2016, McMahon2016, Kako2020, Honjo2021, Reifenstein2021}. The spatial photonic Ising machine is a second example, in which case the spin variables and the spin-coupling are encoded by using spatial light modulators (SLM) while an electrical feedback signal is used to update the spin configuration \cite{Pierangeli2019, Pierangeli2020}. As a final example, we developed an IM based on opto-electronic oscillators (OEO-IM) by using a Mach-Zhender modulator (MZM) to implement the spin-variables and an FPGA to facilitate the spin-coupling and calculate the spin-feedback \cite{Fabian2019, Fabian2021, Fabian2022}. Although the different setups have little in common in terms of used components, they all require an optical modulator, e.g., the aforementioned SLM or MZM. Unfortunately, most commercially available optical modulators lack high modulation resolution, typically offering only around 8bit of resolution \cite{Ehrlichman2008, Yuan2023, Koresawa2020}. Consequently, the feedback signal  in Fig.~\ref{fig: network_example} cannot be represented with arbitrary accuracy. This limited resolution can therefore be seen as a distortion on the feedback signal itself. 

\par
This paper investigates the influence of the resolution of the feedback signal on the performance of the optical IM by making use of numerical simulations. In Section~\ref{sec: Analog IM} and ~\ref{sec: Discretization of the feedback term} we will describe the dynamics of the IM and how the feedback-signal is digitized for different bit-resolutions. Thereafter, in Section \ref{sec: Benchmarking discretization}, we discuss the minimum required bit-resolution of the optical modulators and finally in Section~\ref{sec: 1bit-feedback hardware}, we compare the performance of an IM with a continious feedback signal with that of an IM with a 1bit-resolution optical modulator.

\section{Analog IM}  \label{sec: Analog IM}

\noindent Although the Ising model in Eq.~(\ref{eq: ising_energy}) requires binary spins, many optical hardware implementations make use of analog spin amplitudes $x_i$ \cite{McMahon2016, Fabian2019}. 
To calculate the network's associated energy, a mapping of these analog spin amplitudes to a binary value is required which can be easily implemented by taking for example the sign-function: $sign$($x_i$) = $\sigma_i$. The dynamic evolution of this type of IM is captured by their corresponding dynamic evolution equations, which describe the temporal behavior of the analog spin variables. The type of non-linearity in the dynamic evolution equation depends on the specific hardware implementation. For example, the aforementioned DOPO pulses of the CIM can be modeled as a third-order polynomial, while the MZM of the OEO-IM is characterized by a $\cos^2$-function \cite{Haribara2016, Fabian2019, Fabian2021}. However, the non-linearity in real-world devices is often influenced by clipping effects. Therefore, in this paper, we will use a tangent-hyperbolic as non-linearity as it typically represents these clipping effects well and has a good performance compared to other types of non-linearities \cite{Fabian2021}. Moreover, the sigmoid dynamic evolution equation is already widely employed in neuromorphic systems to emulate the activation patterns of neurons and they can be efficiently implemented in the optical domain \cite{Hopfield1985, Miscuglio2018, Mourgias2019, Jha2020}. The dynamic evolution of the analog spins amplitudes $x_i$ in such an IM can be modeled by:  

\begin{equation}
\label{eq: transfer_function}
    \frac{dx_i}{dt} = -x_i + \tanh(\alpha x_i + \beta\sum_j J_{ij}x_j + \gamma y_i)
\end{equation}
where $\alpha$ represents the gain parameter, $\beta$ the mutual coupling strength, $y_i$ models the colored noise that is inherently present in analog hardware and $\gamma$ the noise strength. The time $t$ is expressed in dimensionless units. As we will see later, the value of $\alpha$ and $\beta$ plays a crucial role in the performance of the IM since these two parameters determine the value that is fed back into the analog hardware.  In this paper, we use a fixed value of 3 for the noise strength $\gamma$ as it is sufficiently high to introduce randomness, yet low enough to avoid noise-dominated dynamics. From now on, we refer to the argument of the hyperbolic tangent, minus the colored noise, as the feedback-term.

The colored noise is described by an extra differential equation:
\begin{equation} 
\label{eq: noise_function}
    \frac{dy_i}{dt} = - \frac{y_i}{\tau} + \zeta_i
\end{equation}
which essentially serves as a low-bandpass filter and is necessary to correctly integrate the noise inside a non-linearity. In Eq.~(\ref{eq: noise_function}), the filter constant $\tau$ is set equal to 0.01 and the Guassian white noise $\zeta_i$ is set to have a mean of 0 and a standard deviation of 1. Note that the noise evolves independent for each spin. 

\par
We simulate the spin dynamics of the analog IM by numerically integrating Eq.~(\ref{eq: transfer_function})--(\ref{eq: noise_function}) using the Milstein method.
The associated energy evolution is obtained by taking the sign of the spin amplitudes after each iteration and substituting this into Eq.~(\ref{eq: ising_energy}). 

\section{Digitization of the feedback term}  \label{sec: Discretization of the feedback term}

\noindent Using numerical simulations, we investigate the influence of the modulation resolution on the performance of IMs by digitization the aformentioned feedback-term. To be able to correctly interpret the results from these simulations, it is crucial to understand how the feedback-term is calculated and how it is digitized. 

First of all, we want to emphasize that in our simulations, we assume the feedback term is calculated correctly, and that a distortion on the signal occurs before it is injected into the non-linearity. The process is visualized in Fig.~\ref{fig: network_example}, where the box with `digitization' is placed after the `feedback' stage to illustrate the sequence of steps. After the digitization of the feedback-term, the noise is added and the signal is sent to the non-linearity. To calculate the feedback-term, the analog spin-amplitudes are needed as well as the coupling matrix \textbf{J}. In an ideal hardware setup, these variables are also implemented in the analog-domain. In our simulations, we use a floating-point number to approximate these analog variables.

\par
Once the feedback-term is calculated, the next step is to account for the finite resolution of the optical modulator in a hardware setup. To do this, the feedback-signal is digitized with a certain bit resolution before being inserted into the hyperbolic tangent function. We refer to this process as the digitization process, where the use of a higher bit resolution results in finer distinctions between the possible output values and thereby enhancing the accuracy of the digitized feedback. We will use the term 'float feedback' to refer to the simulation data obtained using a non-digitized feedback term, and 'bit feedback' for the simulation data obtained with a digitized feedback term. The digitization process itself uses a predetermined digitization interval that is symmetric around zero. This interval is subdivided into equally separated bins and if the input value lies within this interval, it is rounded to its nearest bin value. Otherwise, if the input value lies outside the digitization interval, it is rounded to the closest edge value. The width of the bins is determined by the used bit-resolution since we keep the width of the digitization interval itself constant throughout the simulations. In our simulations, the digitization interval is set to [-4, 4] to ensure that the full output range of the hyperbolic tangent function is utilized. Choosing this interval allows us to cover the entire range of possible output values. Employing a narrower interval would result in a part of the function's output range being inaccessible. The bit-resolution is varied from 1 to 14bit, thereby encompassing a range from 2 to 16,384 bins.

\section{Benchmarking} \label{sec: Benchmarking discretization}

\begin{figure*}[t]
    \begin{center}
        \begin{minipage}[b]{0.60\textwidth}
            \centering
            \begin{minipage}[b]{\textwidth}
                \centering
                \subcaptionbox{\label{fig: g02_float_spinevolution}}{\includegraphics[height=0.30\wdImg, width=\wdImg]{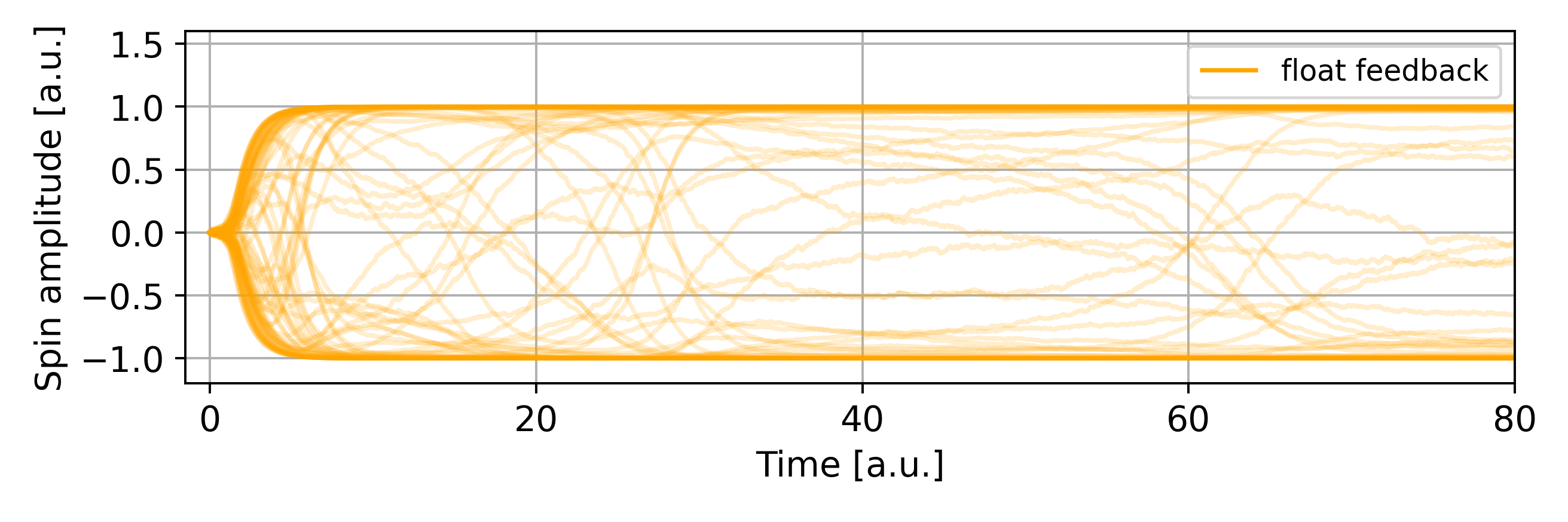}}
            \end{minipage}
	           
            \begin{minipage}[b]{\textwidth}
                \centering
                \subcaptionbox{\label{fig: g02_float_energyevolution}}{\includegraphics[height=0.30\wdImg, width=\wdImg]{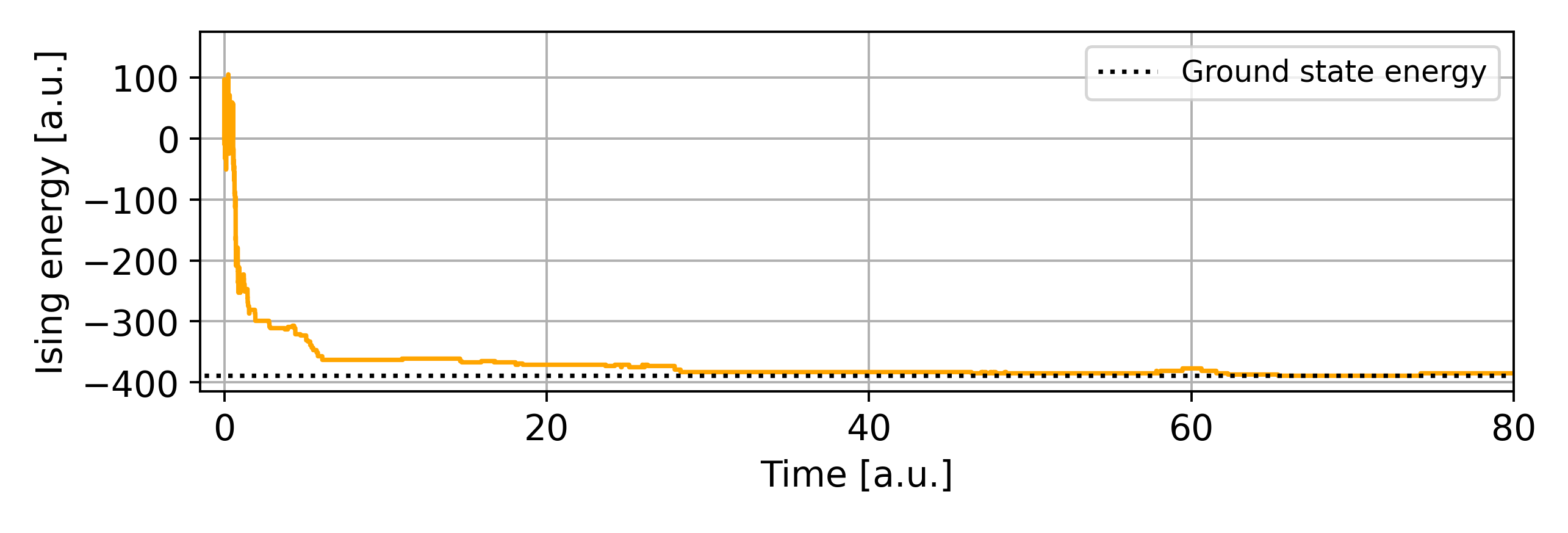}}
            \end{minipage}
        \end{minipage}
        \begin{minipage}[b]{0.39\textwidth}
            \centering
            \subcaptionbox{\label{fig: g02_bitscan_trasnientscr}}{\includegraphics[height=1.0\wdImg, width=\wdImg]{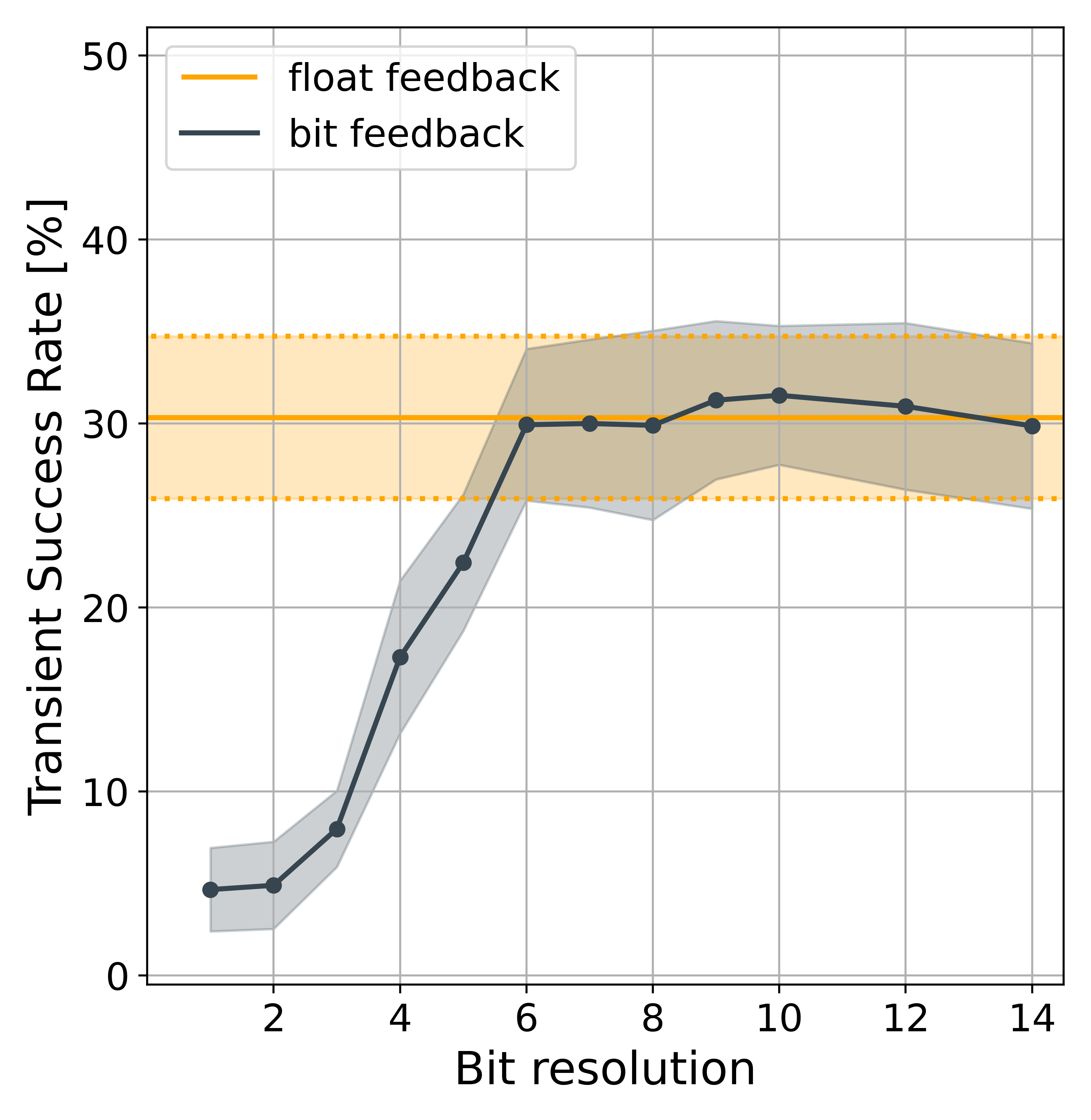}}
        \end{minipage}
    \end{center}
    \caption{ \textbf{(a)} The spin evolution and \textbf{(b)} the energy evolution of the IM solving g05$\_$100.2 biqmac problem with float-feedback. The horizontal line in \textbf{(b)} 	indicates the known ground state energy of the benchmark problem. \textbf{(c)} Transient success rate of  the g05$\_$100.2 biqmac problem, as a function of bit-resolution. The orange and black line indicate the average the transient success rate of the float- and bit feedback respectively. The standard deviation is indicated by the shaded areas around this average.}
    \label{fig: benchmark_discretization}
\end{figure*}

\noindent To determine the required minimum bit-resolution, we run a number of MaxCut benchmark problems from the freely accessible BiqMac and Gset library. MaxCut is a 
NP-hard problem where the objective is to maximize the cut number C by seperatig a graph into two parts \cite{Garey1979, Lucas2014}. The cost function of this optimization problem is given by:
\begin{equation}
\label{eq: cut_value}
    C = \frac{1}{4} \left( \sum_{ij} J_{ij} - \sum^{N}_{ij} J_{ij} \sigma_i  \sigma_j    \right)
\end{equation}
and maximizing this cost function is equivalent to minimizing the Ising energy given by Eq.~($\ref{eq: ising_energy}$). From the BiqMac library, we specifically use the MaxCut ''g05$\_N.x$''-problem sets, where $N$ indicates the network size and $x$ the number of the problem at hand. The size of these problem sets is 60, 80 or 100 spins and all of them have an average edge-probability of 50$\%$. The Gset library is used to access benchmark problems with larger network sizes of 800 and 1000 spins. 

\par
Fig.~\ref{fig: benchmark_discretization}(a) shows the evolution of all 100 spins when solving the "g05\_100.2" benchmark problem with the conventional float feedback. To avoid any transient behavior of the noise, we use a warm-up of 0.5 time units before instantaneously changing $\alpha$ and $\beta$ from zero to their optimum value. From that moment on, the spin variables start to evolve following the dynamics of Eq.~(\ref{eq: transfer_function}), and after approximately 65 time units the IM reaches the ground-state energy. This is seen more clearly in Fig.~\ref{fig: benchmark_discretization}(b), which visualizes the associated energy evolution of the system and where the horizontal dashed line indicates the ground-state energy. However, the IM will not always reach the ground-state energy because it can get stuck in local energy minima. Therefore, the performance of the IM is typically characterized by the probability of finding the ground-state energy during its runtime, the so-called transient success rate (TSR). The TSR is thus defined as the ratio of the number of runs wherein the IM reaches the ground-state energy over the total number of runs. We will use the TSR as a metric to assess the influence of the bit-digitization on the performance of the IM. The TSR is extracted from 100 independent runs for each considered benchmark problem and each bit resolution. Afterwards, we check for which bit resolution the TSR of the bit- and float feedback simulations start to overlap and consider this the minimum required bit resolution.

\par
For every problem, we have first optimized the value of $\alpha$ and $\beta$ with respect to the TSR for an IM with float feedback using a brute force parameter scan. Although obtaining these optimal hyperparameters is certainly usefull, they do not influence the conclusion of this paper. 

\par
We calculate for each benchmark problem the TSR 30 times, from which we subsequently determine the average and the spread of the TSR distribution. As a representative example, the bit-resolution scan for the ''g05$\_$100.2'' benchmark problem is shown in Fig.~\ref{fig: benchmark_discretization}(c). The horizontal orange line corresponds to the average TSR for the IM with float feedback, with the dotted lines indicating the standard deviation of the 30 simulations around the average value. The average TSR for the IM with the bit feedback is indicated in black. In this case, the average TSR values are represented as data points, while the shaded area around them illustrates the standard deviation of the TSR distribution for that particular bit-resolution. From this figure, it is evident that the TSR of the IM with bit feedback increases with bit-resolution up to 6bit. At this point, the average TSR of the bit feedback overlaps with the one-standard deviation range of the TSR for the float feedback. Moreover, further increasing the bit-resolution does not result in a further increase of the TSR. This means that for this specific benchmark problem, a 6bit-resolution is sufficient for the hardware implementation of the modulator.

\begin{figure*}
    \begin{center}
        \begin{minipage}[b]{0.59\textwidth}
            \centering
            \subcaptionbox{\label{fig: min_bit_res}}{\includegraphics[width=\textwidth]{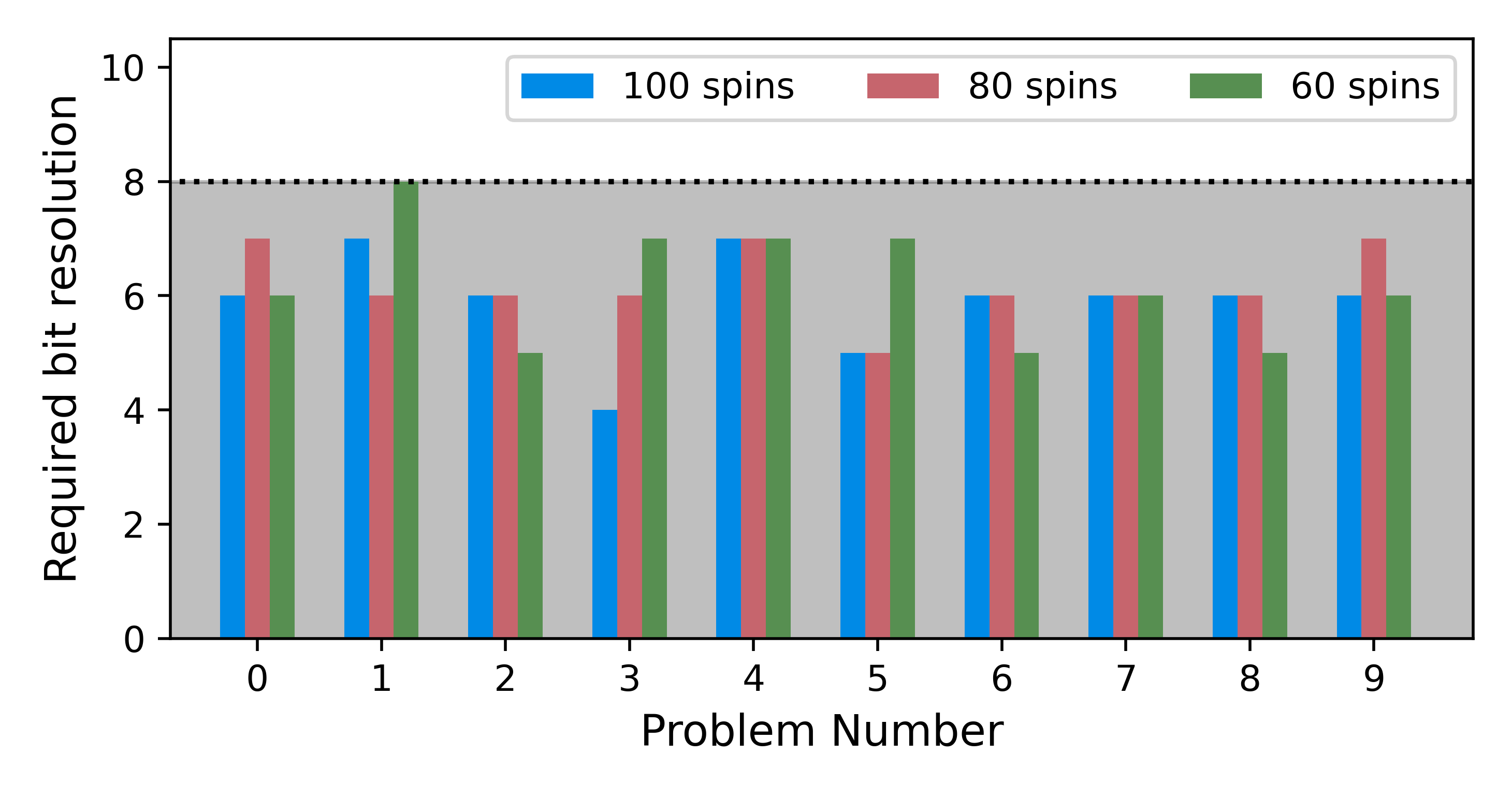}}
        \end{minipage}
        \hfill
        \begin{minipage}[b]{0.40\textwidth}
            \centering
            \subcaptionbox{\label{fig: min_bit_scaling}}{\includegraphics[width=\textwidth]{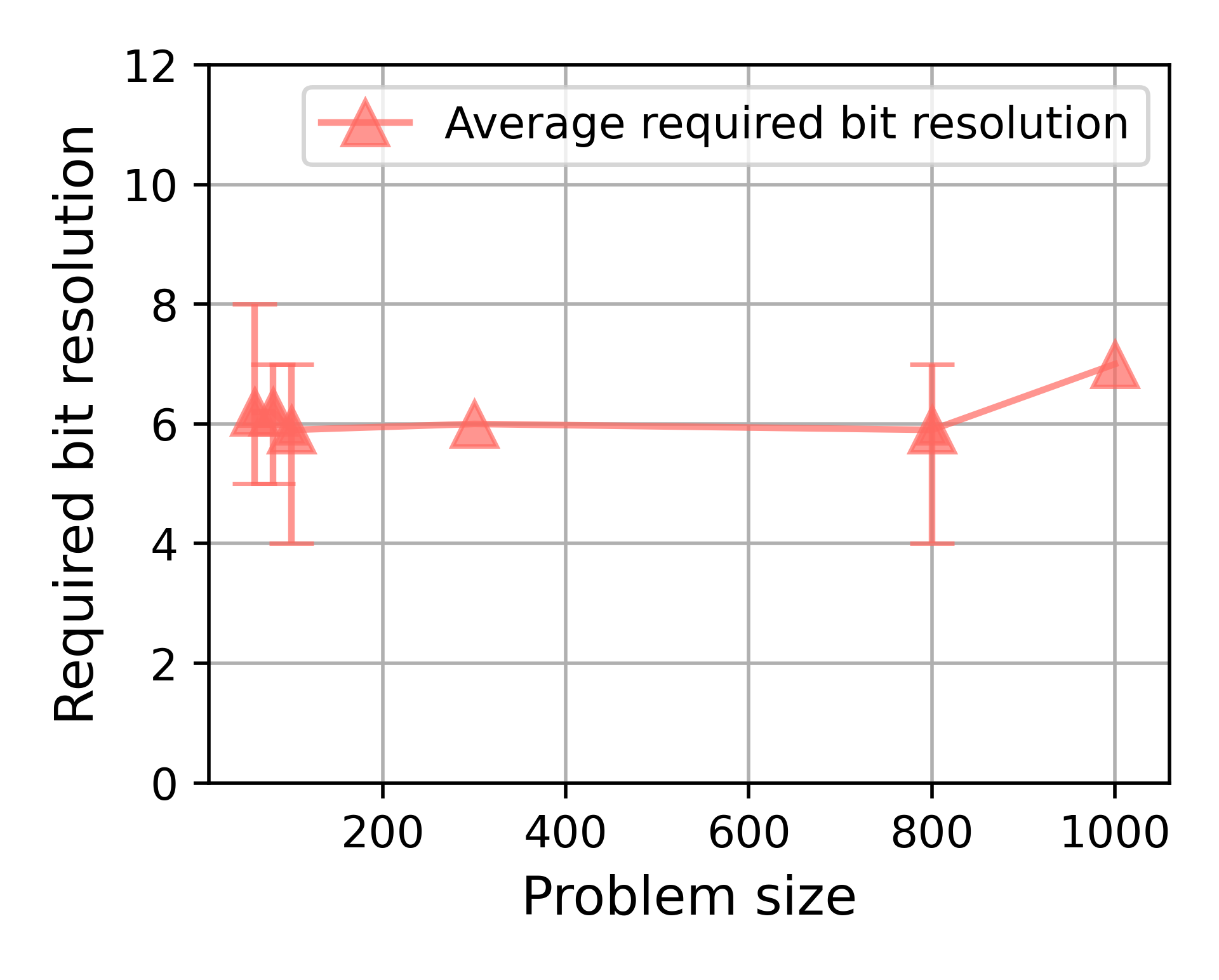}}
        \end{minipage}
    \end{center}
    \caption{\textbf{(a)} The required bit resolution for three different problem sets is evaluated, with each set consisting of 10 MaxCut problems. Each problem set has a different size (60, 80, and 100 spins) and is represented by a distinct color. The x-axis shows the problem numbers for each set. A horizontal dashed line at 8bit indicates the minimum required bit resolution for all the benchmark problems considered. \textbf{(b)} The required bit resolution for different problem sizes. For the network size uptill 300 spins, the Biqmac library is used, while for larger problem sizes the Gset library was used. The red triangle visualizes the average required bit-resolution for each problem size while te error bars obtained bit-resolution in each used data set.}
    
    \label{fig: problem_digitzization}
\end{figure*}

\par
Similar simulations have been performed on the other BiqMac and Gset benchmark problems. The resulting minimum required bit-resolutions for the BiqMac benchmark problems are summarized in Fig.~\ref{fig: problem_digitzization}(a), where the color signals the problems size, the horizontal axis shows the problem number and the vertical axis the associated minimum required bit-resolution of that specific problem. From this figure, it is clear that there exists a spread on the minimum required bit resolution. For example, for the ''g05$\_$100.3'' benchmark problem, only a 4bit resolution is needed. On the other hand, the ''g05$\_$60.2'' has the highest minimum required resolution with 8bit, making it the overall minimum required bit-resolution to accurately represent the feedback-term in these benchmark problems. 

\par
To check whether the problem size influences the minimum required bit resolution, we use larger benchmark problem taken from \cite{Shi2024} and from the Gset library. For these problems, we do not make use of the TSR as the success probability on these benchmark problems is too low. We therefore use the TSR98, which is defined as the probability of reaching 98\% of the best-known cut value, defined in Eq.~(\ref{eq: cut_value}). The other simulation steps are the same as before. Fig.~\ref{fig: problem_digitzization}(b) summarizes the results, where on the x-axis the problem size is depicted and on the y-axis the minimum required bit resolution is shown. The red triangle visualizes the average required bit-resolution, based on the distribution of each used data set, while the vertical error bars are used to indicate the max- and minimum obtained bit-resolution in each used data set. Note that the data points for the problem size up to 60, 80 and 100 spins are from the BiqMac library, which are also shown in Fig.~\ref{fig: problem_digitzization}(a). The ten problems involving 300 spins are taken from \cite{Shi2024} but have the same characteristics as the ones from the BiqMac library. From the Gset library, we have used problems G1-G10 for the 800-spin problems and the G43-G47, G51-G54 for the 1000-spin problems. The data points of the problem set with 300 spins and 1000 spins are all overlapping at 6- and 7bit respectively, hence there is no spread visible at these points. The fact that these data points overlap indicates that the problems selected within these sets are all rather similar. Overall, the required bit resolution does not seem to strongly depend on problem size and 8bits of resolution is sufficient for all investigated problems. The increasing time and memory requirements for larger problem sizes limited our investigation of even bigger problem sets.

\par 
Overall, we can conclude from Fig.~\ref{fig: problem_digitzization}(b) that the IM implementation should have at least an 8bit modulator to correctly emulate the float-IM of the simulations which is achievable with many modern-day optical components \cite{Ehrlichman2008, Yuan2023, Koresawa2020}. The use of optical components with higher resolution modulators does not seem to provide any additional benefits to the performance of the IM. We believe that these results can help optimize the cost vs. performance of future IM setups.

\section{1bit-feedback hardware} \label{sec: 1bit-feedback hardware}

\begin{figure*}
    \begin{center}
        \begin{tabularx}{\textwidth}{*{3}{s}}
            \subcaptionbox{\label{fig: subfig_scr_2}}{\includegraphics[width=\wdImg]{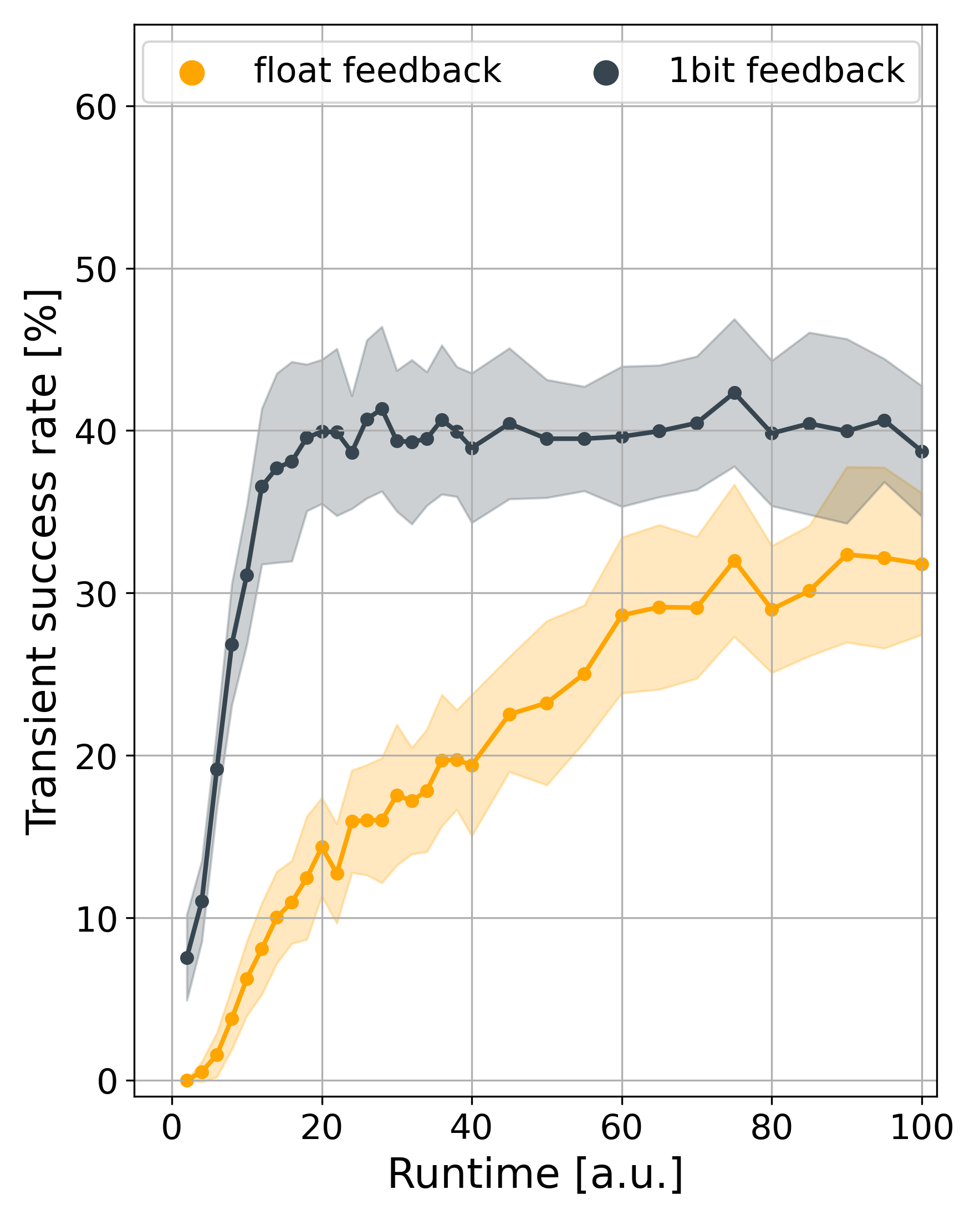}}
            & \subcaptionbox{\label{fig: subfig_scr_4}}{\includegraphics[width=\wdImg]{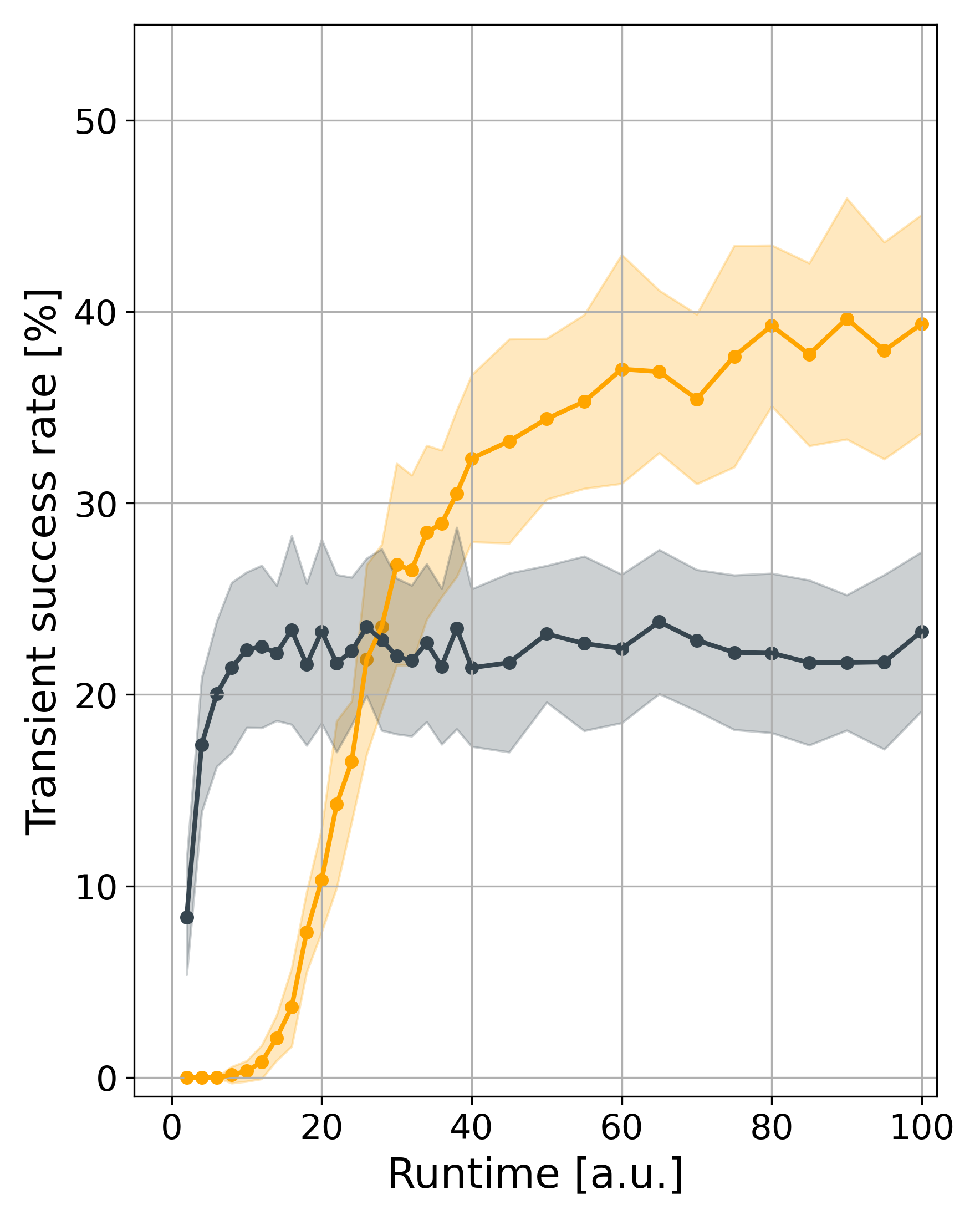}}
            & \subcaptionbox{\label{fig: subfig_scr_9}}{\includegraphics[width=\wdImg]{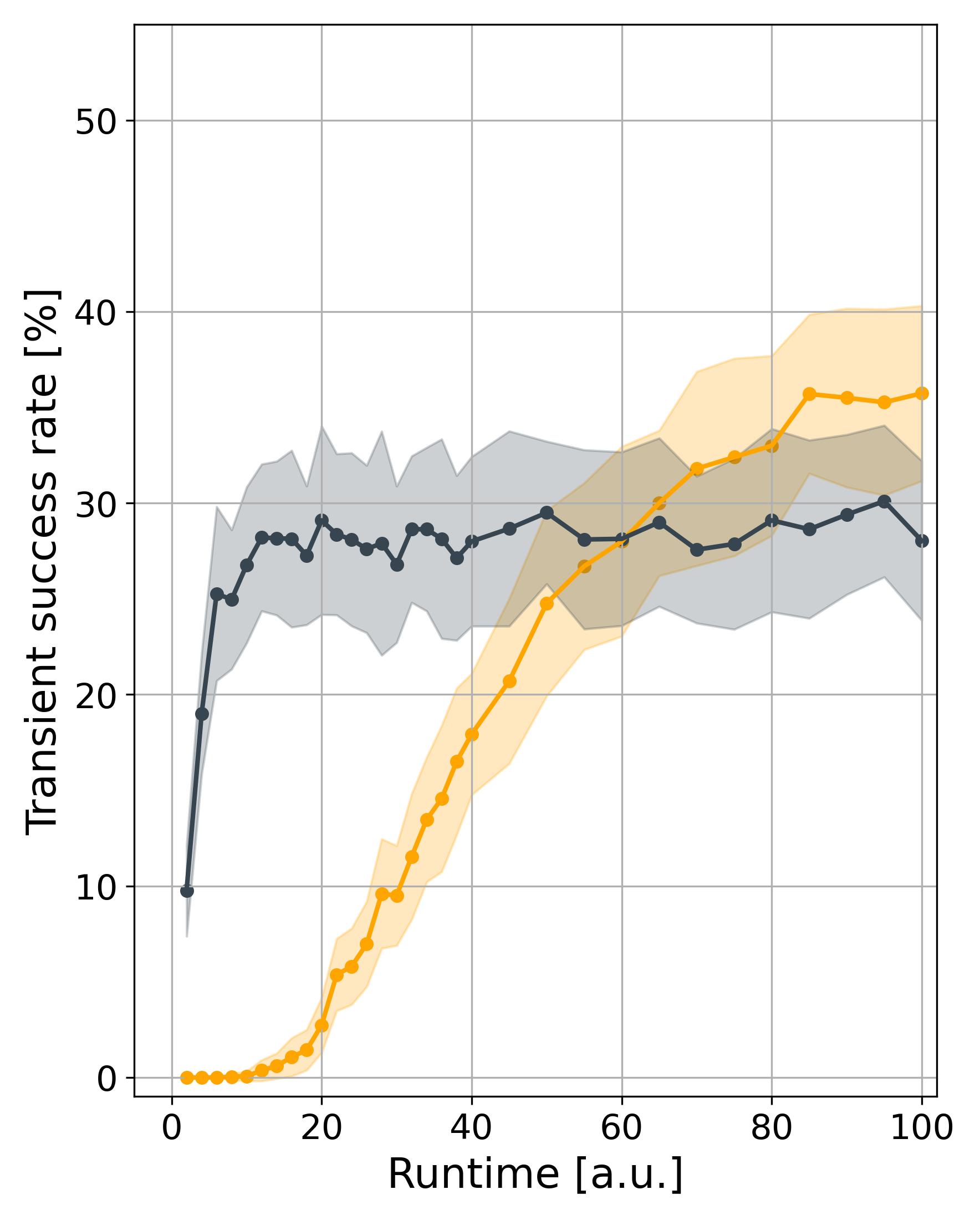}}
            \\
            \subcaptionbox{\label{fig: subfig_tts_2}}{\includegraphics[width=\wdImg]{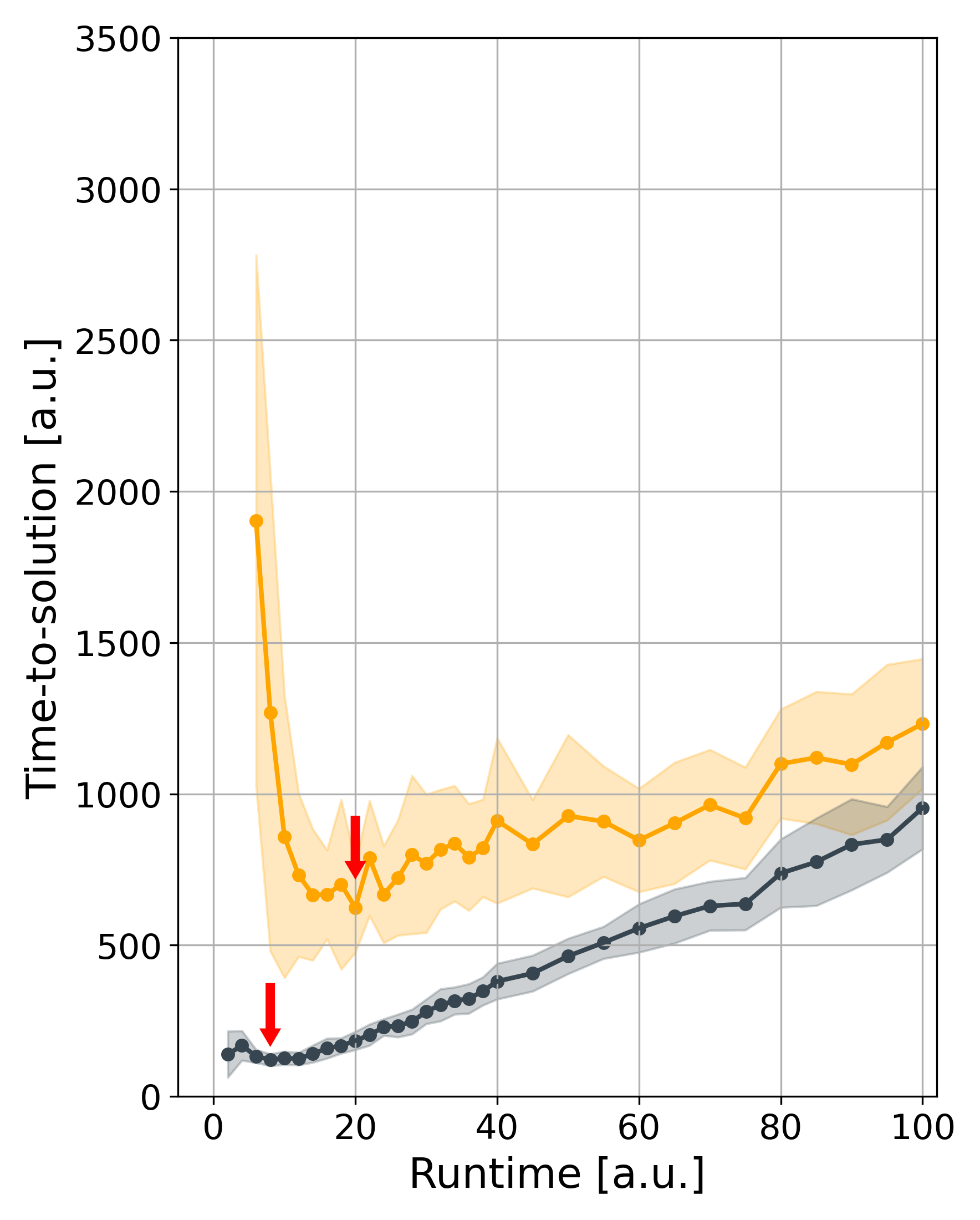}}
            & \subcaptionbox{\label{fig: subfig_tts_4}}{\includegraphics[width=\wdImg]{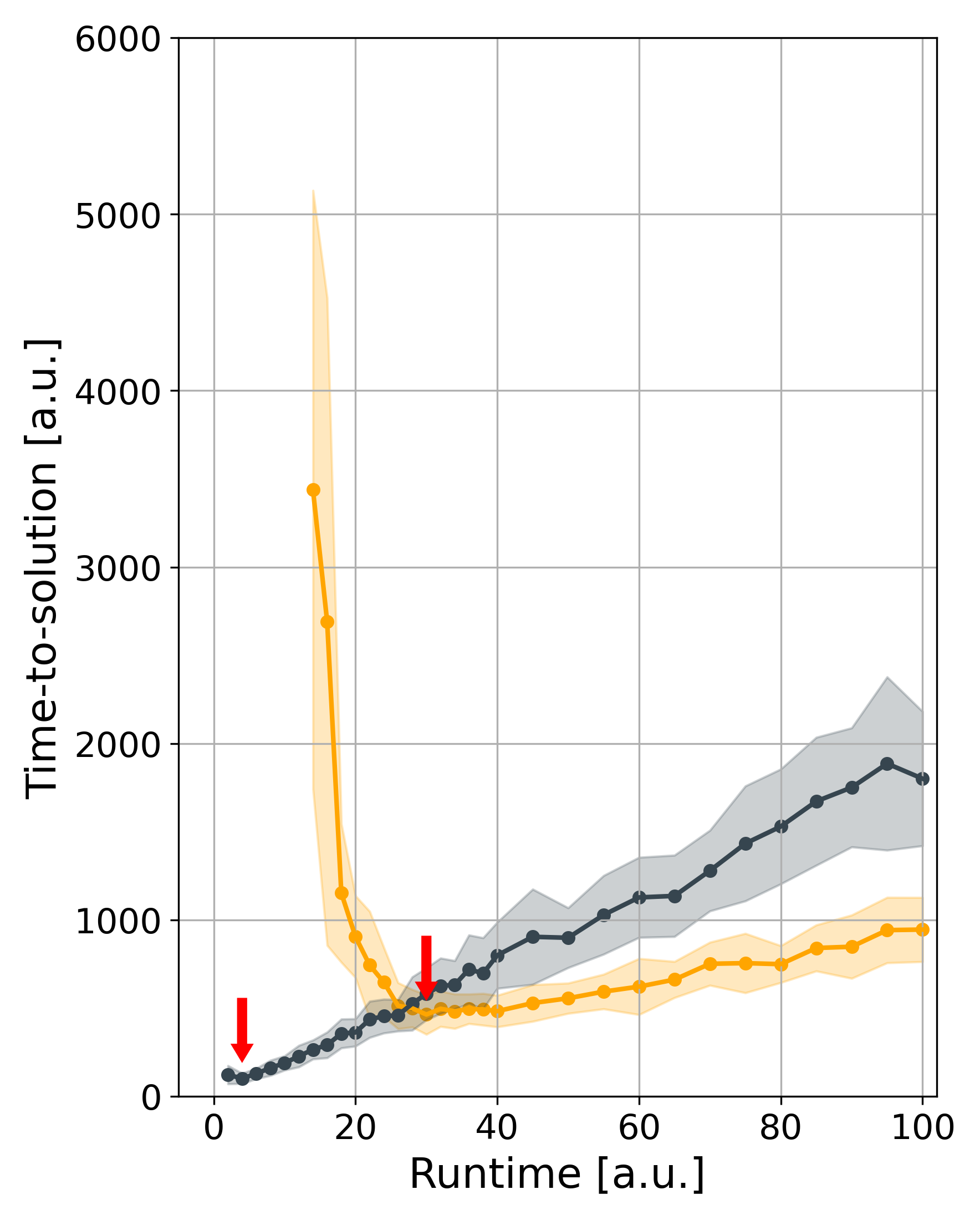}}
            & \subcaptionbox{\label{fig: subfig_tts_9}}{\includegraphics[width=\wdImg]{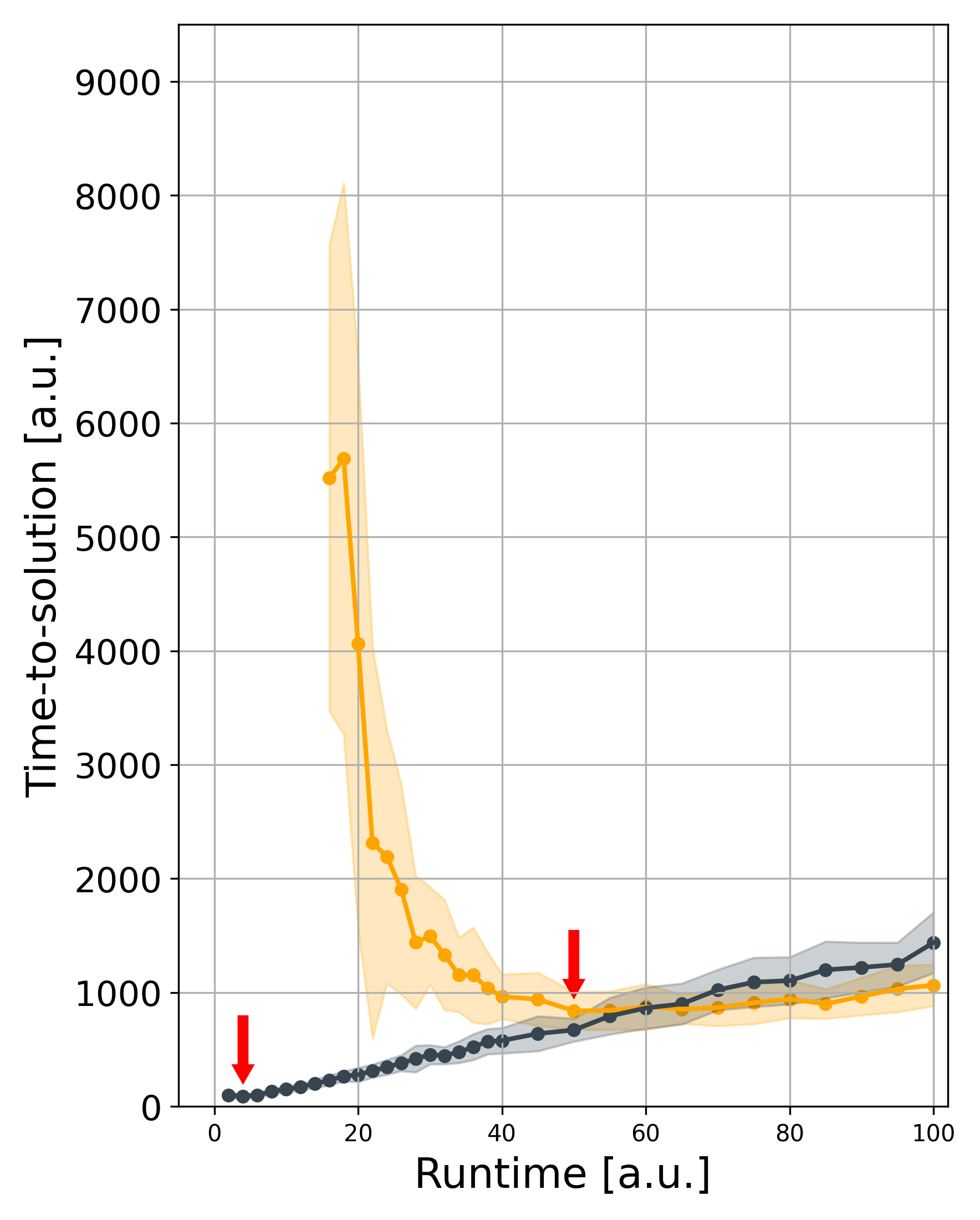}}
            \\
        \end{tabularx}
    \end{center}
    \caption{The average transient success rate and TTS as a function of runtime. \textbf{(a)}-\textbf{(c)} show how the average transient success rate changes with increasing runtime. The considered MaxCut problems are the 100.2, 100.4 and 100.9 respectively. All three problems are taken from the BiqMac library. \textbf{(d)}-\textbf{(f)} show how the average TTS changes with runtime. The problems considered are again 100.2, 100.4 and 100.9 respectively. The red arrowes inidacte the effective TTS and in all three cases, the 1bit feedback system has a smaller effective TTS than the float feedback system.}
    \label{fig: SCR_TTS_runtimescan}
\end{figure*}

\par
From Fig.~\ref{fig: benchmark_discretization}(c), it is interesting to note that resolutions lower than the found minimum required bit-resolution still result in a finite TSR, even for the lowest possible resolution of 1bit. It is therefore interesting to investigate the computational power of such a 1bit feedback hardware as it has the important advantage of a lower power consumption \cite{Nahmias2019_1bitenergy}, but also easier design and lower fabrication cost \cite{Liu2019_1bitdesign}.

\par
As mentioned before, the performance of the IM depends on the value of $\alpha$ and $\beta$. In the previous section, we have optimized these hyperparameters with the assumption that the system receives a float feedback. However, as we are comparing now the performance of two seperate systems, i.e. an IM with float feedback and an IM with 1bit feedback, we optimize the two feedback parameter $\alpha$ and $\beta$ for each system separately. Furthermore, to thoroughly compare the performance of the two systems, we use both the average TSR but also the average time-to-solution (TTS) as performance metrics.

\begin{figure*}[b]
    \centering
    
    \begin{minipage}{\textwidth}
        \centering
        \raisebox{0.78cm}{\large (a)}
        \hfill
        \includegraphics[width=0.96\textwidth]{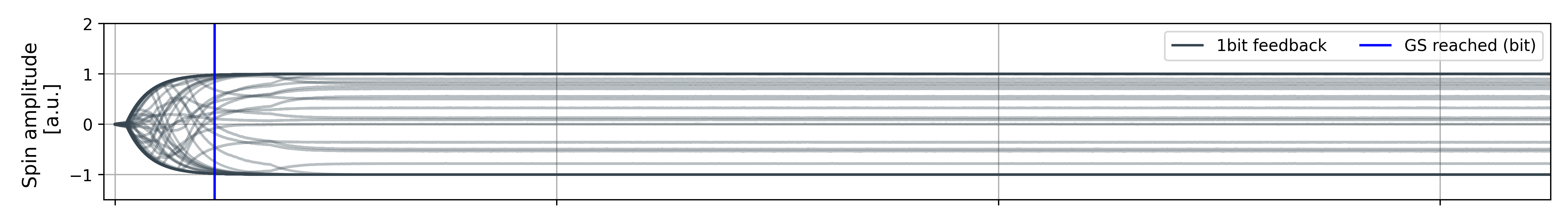} 
    \end{minipage}
    \vspace{0em}
    
    \begin{minipage}{\textwidth}
        \centering
        \raisebox{0.78cm}{\large (b)}
        \hspace{0.0cm}
        \includegraphics[width=0.96\textwidth]{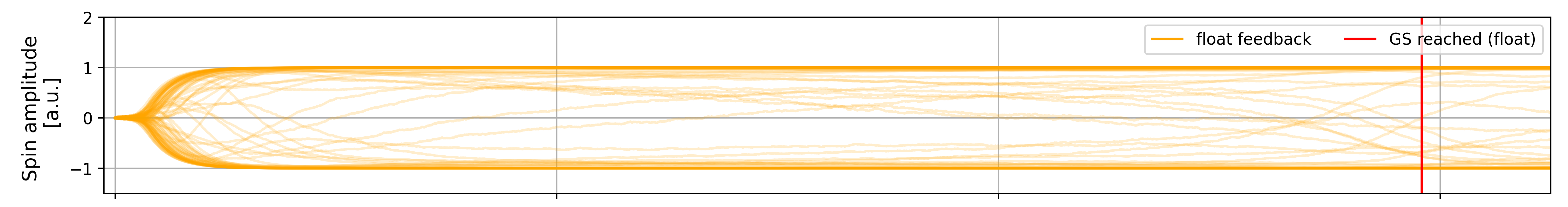}
        \hfill
    \end{minipage}
    
    \vspace{-1.0em}
    
    \begin{minipage}{\textwidth}
        \centering
        \raisebox{1.5cm}{\large (c)}
         \hspace{0.0cm}
        \includegraphics[width=0.96\textwidth]{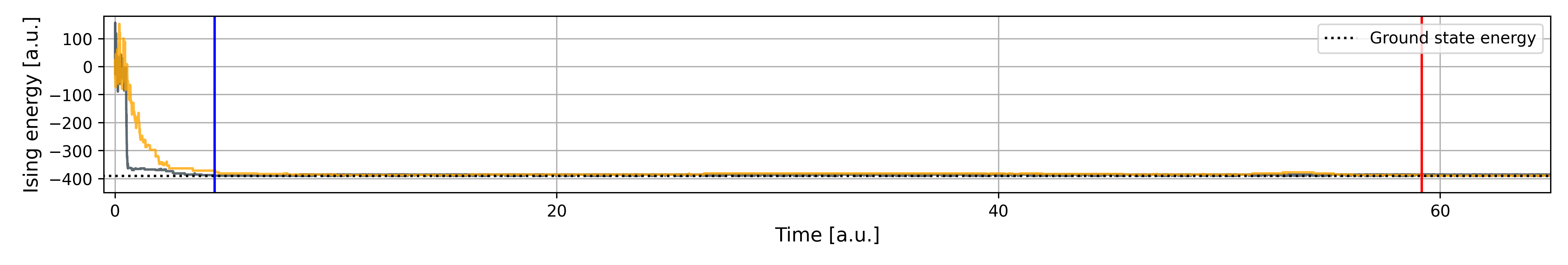}
    \end{minipage}
    
     \vspace{0em}

    \caption{ \textbf{(a)} Spin evolution of the system with the an optimized bit feedback and \textbf{(b)} an optimized float feedback. The vertical blue (red) line indicates the time at which the GS is first reached for the 1bit feedback (float feedback). Pannel \textbf{(c)} shows the associated energy evolution of both feedback systems.}
    
    \label{fig: optimized_evolutions}
\end{figure*}

\begin{table*}[t]
\centering
\caption{Effective TTS value for the considered biqmac problems. The second and third row show the effective TTS value for the float feedback and the 1bit feedback system, respectively. The last row shows the change in effective TTS value, when comparing the 1bit feedback with the float feedback.}
\begin{adjustbox}{width=\textwidth}
\begin{tabular}{|c|c|c|c|c|c|c|c|c|c|c|}
        \hline
        \textbf{Problem} & 100.0 &  100.1 & 100.2 & 100.3 & 100.4 & 100.5 &  100.6 & 100.7 & 100.8 & 100.9 \\
        \hline
       \textbf{float feedback}  &  1345 & 1994 & 624 & 3326 & 464 & 2791 & 472 & 569 & 7292 & 840 \\
        \hline
        \textbf{1bit feedback}  & 222 & 556 & 121 & 2216 & 102 & 252 & 81 & 92 & 1503 & 90 \\
        \hline
        \hline
       \textbf{Relative change [$\%$]}  & -83.5 & -72.1 & -80.7 & -33.4 & -78.0 & -91.0 & -82.9 & -83.8 & -79.4 & -89.3 \\
       \hline    
    \end{tabular}
\end{adjustbox}
\label{tab: min_bit_tts}
\end{table*}

\par
The TTS is a useful and widely used metric when comparing the performance of different heuristic hardware solvers, as it is not only important to consider how often the system finds the solution of the problem at hand, but also the time required to find it. We define the TTS as the time that one would need to run the IM to find the solution with 99\% certainty. Mathematically, this TTS definition is written as
\begin{equation}
    \textrm{TTS} = 
    \begin{cases}
        \textrm{T}_r \frac{\log(0.01)}{\log(1 - \textrm{TSR})} & \textrm{for } 0.0 < \textrm{TSR} \leq 0.99 \\
        \textrm{T}_r & \textrm{for } \textrm{TSR} = 1.0 \\
    \end{cases} 
\label{eq: time-to-solution}
\end{equation}
where T$_r$ is the runtime of the IM. From this definition, there is a clear dependence on T$_r$, making it necessary to calculate TSR as a function of runtime before inserting it into Eq.~(\ref{eq: time-to-solution}) to find the associated TTS. Next, the minima of this runtime scan is taken as the effective TTS of the benchmark problem.

\par
Fig.~\ref{fig: SCR_TTS_runtimescan}(a)--\ref{fig: SCR_TTS_runtimescan}(c) show the average TSR for three separate benchmark problems from the BiqMac library, namely ''g05$\_$100.2'', ''g05$\_$100.4''  and ''g05$\_$100.9'', for both the float- and 1bit feedback as a function of the runtime T$_r$. For each feedback method and benchmark problem, a single data point represents the average TSR taken from a distribution of 30 separately calculated TSR values, each determined from 100 simulation runs. The standard deviation of this distribution is indicated by the shaded area around the average data points. As can be seen from the three benchmark problems that are shown in Fig.~\ref{fig: SCR_TTS_runtimescan}(a)--\ref{fig: SCR_TTS_runtimescan}(c), the average TSR for both float- and 1bit feedback initially increases with runtime, but eventually starts to stagnate. This behavior is anticipated because increasing the runtime of the IM initially provides more time for the system to change the analog spin amplitudes, consequently increasing the probability of finding the ground-state energy. After a certain amount of runtime, however, the system will reach a (meta-)stable configuration, and a further increase in runtime will not enhance the TSR anymore.

\par 
The runtime scan of these three specific benchmark problems is shown to illustrate that the effect of the 1bit feedback on the average TSR varies for different benchmark problems. Using a 1bit feedback can result in a decrease, an increase, or similar average TSR compared to a system with float feedback at T$_r$=100. However, for all the considered benchmark problems, the average TSR in the 1bit feedback simulations stagnates significantly earlier than the simulations with the float feedback. To check whether or not the higher average TSR that is observed in some of the float feedback simulations is worth the longer runtime, we calculate the TTS for both the float- and 1bit feedback for each of the ''g05$\_$100'' BiqMac problems.

\par
Using the 30 independent TSR data points, the associated average TTS data points for both the float- and the 1bit feedback are calculated using Eq.~(\ref{eq: time-to-solution}) and are subsequently depicted in Fig.~\ref{fig: SCR_TTS_runtimescan}(d)--\ref{fig: SCR_TTS_runtimescan}(f). The problems considered are again ''g05$\_$100.2'' , ''g05$\_$100.4''  and ''g05$\_$100.9'' respectively and the spread of the curves are once-more the standard deviation of the 30 TTS-values obtained from simulations at each runtime. At this point it is worth noting that the TTS for the float feedback at small runtimes is not always defined as the TSR goes to zero. To ensure that we had a sufficient number of non-zero TSR data points to maintain a representative distribution, we set a threshold requiring at least 60$\%$ of the TSR data points to be non-zero. If this threshold was not met, the TTS was considered undefined, and these data points were excluded from the figures. 

The qualitative trend observed in Fig.~\ref{fig: SCR_TTS_runtimescan}(d)--\ref{fig: SCR_TTS_runtimescan}(f) is also representative for all the considered benchmark problems. First, the TTS decreases with increasing runtime $T_r$ for both the float and the 1bit feedback before reaching a minimum value, indicated by a red arrow. For still larger runtime values, the TTS increases again monotonically. This behavior can be understood from the runtime scans of the TSR, which were explained previously. Initially, the steep increase in TSR results in a decrease of the TTS, but as the average TSR starts to stagnate for longer runtimes, its increase can no longer compensate for the increasing runtime, and hence the TTS increases. As mentioned before, it is the minimum of these curves that is taken as the effective TTS of the benchmark problem, which is always lower for the 1bit feedback, not only for the three examples shown but for all the 10 considered benchmark problems. The exact values of the effective TTS for both the float and the 1bit feedback are summarized in Tab.~\ref{tab: min_bit_tts}, along with the relative change in effective TTS value when comparing the 1bit feedback with the float feedback. From the table, one can see that the 1bit feedback system reduces the minimum average TTS by an average of 77.4\%.

\par
These numerical results show that using a feedback signal with a resolution of 1bit will speed up the ground-state search of the IM by an order of magnitude for almost all the considered benchmark problems. While the average TSR of an IM with 1bit feedback may be lower than that of an IM with float feedback, it is possible for the 1bit feedback system to complete multiple runs within the same time span as the float feedback IM. To illustrate this idea, Fig.~\ref{fig: optimized_evolutions} compares the spin and energy evolution of the optimized float and 1bit feedback systems for the "g05\_100.2" benchmark problem. The red and blue vertical lines represent the time at which the bit feedback and the float feedback reach the ground state energy, respectively. Looking at the spin evolutions in Fig.~\ref{fig: optimized_evolutions}(a)--\ref{fig: optimized_evolutions}(b), the spin amplitudes of the 1bit feedback bifurcates more abruptly due to the relatively high feedback signal that its spins receive, after which it reaches the ground state energy around 4.5 time units. In contrast, the spin amplitudes that receive the float feedback evolve significantly slower and only reach the ground state energy after about 59 time units. In the same amount of time, the 1bit feedback system could already have been run 7 times since its spins are stable after 10 time units, hence increasing its overall probability of reaching the ground state. The slower spin dynamics are also observed in the associated energy evolution, shown in Fig.~\ref{fig: optimized_evolutions}(c), where the energy of the 1bit feedback system drops significantly faster compared to the float feedback system. 

\par
Finally, we focused in this paper on the impact of the feedback term's bit resolution on performance. Other optical components in the IM can lead to digitization of the signal at other points in Fig.\ref{fig: network_example}, such as when implementing the coupling weights. The effects of these additional digitization steps are beyond the scope of this paper and will be investigated in future work.

\section{Conclusion} \label{sec: Conclusion}
In this paper, we have explored the effects of the bit resolution of the feedback term on the performance of an IM. From numerical simulations, we found that a resolution of 8bit is sufficient to obtain the same TSR as when using a non-digitized feedback. Moreover, by using problem sizes as big as 1000 spins, we observe no clear scaling of this minimum required bit resolution with the problem size.

\par
Secondly, we observed that the IM still has a finite TSR for feedback resolutions as low as 1 bit. Although the average TSR of the 1bit feedback system is often lower, it has a much lower TTS compared to the float feedback system due to its rapid spin stabilization. This decrease, which is of an order of magnitude for almost all the benchmark problems, results in a significant improvement in performance that could be leveraged to enhance future IM systems. Moreover, the use of such a 1bit-modulator can be expected to decrease the enery consumption and cost of the IM.

\section{Data availability} \label{sec:  Data availability}
\noindent The authors declare that all relevant data are included in the manuscript. Additional data are available from the corresponding author upon reasonable request.

\section{Author contributions} \label{sec: Author contributions}
\noindent T.S. performed the simulations and wrote the manuscript. G.V.d.S and G.V. supervised the project. All authors discussed the results and reviewed the manuscript.

\section{Additional information} \label{sec:  Additional information}
\noindent \textbf{Competing interest: }The authors declare no competing interests. 
\newline 	
\noindent  \textbf{Acknowledgements: }The authors would like to thank Thomas Van Vaerenbergh for the insightful discussions. This research was funded by the Research Foundation Flanders (FWO) under grants G028618N, G029519N and G006020N. Additional funding was provided by the EOS project ”Photonic Ising Machines”. This project (EOS number 40007536) has received funding from the FWO and F.R.S.-FNRS under the Excellence of Science (EOS) programme.

\bibliographystyle{IEEEtran}
\bibliography{IEEEabrv, BitResolution_texfile}

\end{document}